\documentclass[twocolumn,prb,showpacs]{revtex4}
\usepackage{amsmath,amsthm}
\usepackage{epsfig}
\usepackage{graphics}
\usepackage{graphicx}
\usepackage{dcolumn}
\usepackage{bm}

\begin{document}

\title{Infrared spectroscopy and nano-imaging of the insulator-to-metal transition in vanadium dioxide}
\author{M. M. Qazilbash,$^{1,\ast}$ M. Brehm,$^2$ G. O. Andreev,$^1$ A. Frenzel,$^1$ P.-C. Ho,$^{1,3}$ Byung-Gyu Chae,$^4$
 Bong-Jun Kim,$^4$ Sun Jin Yun,$^4$ Hyun-Tak Kim,$^4$, A. V.
Balatsky,$^5$ O. G. Shpyrko,$^1$ M. B. Maple,$^1$ F. Keilmann,$^2$
and D. N. Basov$^1$}
\affiliation{$^{1}$Physics Department, University of California-San Diego, La Jolla, California 92093, USA. \\
$^{2}$Abt. Molekulare Strukturbiologie, Max-Planck-Institut
f$\ddot{u}$r Biochemie and Center for NanoScience,
82152 Martinsried, M$\ddot{u}$nchen, Germany. \\
$^{3}$Department of Physics, California State University-Fresno,
Fresno, California 93740, USA. \\
$^{4}$IT Convergence and Components Lab, Electronics and
Telecommunications Research Institute (ETRI), Daejeon 305-350, Korea. \\
$^{5}$Theoretical Division and Center for Integrated
Nanotechnologies, MS B262, Los Alamos National Laboratory, Los
Alamos, New Mexico 87545, USA.}

\date{\today}

\begin{abstract}

We present a detailed infrared study of the insulator-to-metal
transition (IMT) in vanadium dioxide (VO$_2$) thin films.
Conventional infrared spectroscopy was employed to investigate the
IMT in the far-field. Scanning near-field infrared microscopy
directly revealed the percolative IMT with increasing temperature.
We confirmed that the phase transition is also percolative with
cooling across the IMT. We present extensive near-field infrared
images of phase coexistence in the IMT regime in VO$_2$. We find
that the coexisting insulating and metallic regions at a fixed
temperature are static  on the time scale of our measurements. A
novel approach for analyzing the far-field and near-field infrared
data within the Bruggeman effective medium theory was employed to
extract the optical constants of the incipient metallic puddles at
the onset of the IMT. We found divergent effective carrier mass in
the metallic puddles that demonstrates the importance of
electronic correlations to the IMT in VO$_2$. We employ the
extended dipole model for a quantitative analysis of the observed
near-field infrared amplitude contrast and compare the results
with those obtained with the basic dipole model.

\end{abstract}


\maketitle

\section{Introduction}

Vanadium dioxide (VO$_2$) undergoes an insulator-to-metal
transition (IMT) at $T_c$ $\approx$ 340 K. The IMT is accompanied
by a change of lattice structure from a monoclinic ($M$1) unit
cell in the low-temperature insulating phase to tetragonal
(rutile) symmetry in the metallic phase. VO$_2$ has been studied
extensively for two main reasons. First, the driving mechanism of
the IMT has remained controversial and many experimental and
theoretical studies have focused on understanding
it.\cite{mottbook,imada-rmp} Secondly, VO$_2$ has immense
potential for applications \cite{apllopez,aplmmq,apltom} because
the phase transition occurs only 40 K above room temperature and
involves dramatic changes in transport, infrared and optical
properties. Moreover, the phase transition can also be triggered
optically \cite{optlettrini,aplrini} and electrically
\cite{aplmmq,aplkim} leading to new optical and electronic
devices.

The reason the driving mechanism of the IMT in VO$_2$ has been a
subject of controversy is the complex nature of the interactions
in this material. VO$_2$ consists of just two elements, vanadium
and oxygen, with well-documented lattice structures in the M1 and
rutile phases. The model of delocalized \emph{d}-electrons of
vanadium (one per vanadium ion) moving in the periodic potential
created by the lattice was expected to explain the documented
features of the IMT in VO$_2$. In the early literature, the phase
transition in VO$_2$ was described in terms of the structural
distortion that leads to the formation of vanadium pairs and
unit-cell doubling.\cite{goodenough} This Peierls instability was
thought to occur due to the coupling of the conduction electrons
to a soft phonon mode of vanadium.\cite{hearn,gupta,diffuseXray}
However, the model of a Peierls instability within band theory
alone ~\cite{allenprl,eyert} cannot satisfactorily explain several
experimental aspects of the phase transition and the properties of
the insulating and metallic phases existing in its vicinity.
Following are a few examples of such experimental observations:
the $\approx$ 0.6 eV energy gap in the $M$1 insulator measured by
infrared and photoemission
techniques~\cite{barkerprb,shin,okazaki,VO2mmq1,VO2mmq2}; the
changes in optical conductivity and spectral weight up to at least
6 eV due to the IMT~\cite{barkerprb,shin,okazaki,VO2mmq1,VO2mmq2};
the satellite feature at $\approx$ 1 eV below the Fermi energy
seen in photoemission in the rutile metal which may well be a
remnant of the lower Hubbard band~\cite{xray-photo}; and lack of
resistivity saturation in the rutile metal beyond the
Ioffe-Regel-Mott limit of metallic transport which is difficult to
explain within a framework based on electron-phonon
coupling.\cite{VO2mmq1,allenprb,kivelson} There is growing
evidence that electron-electron correlations cannot be neglected
in any viable description of the IMT in
VO$_2$.\cite{VO2mmq1,VO2mmq2,mmqscience,kimnewjphys,kimprl,lupi,kimprb}
Several previous theoretical studies have emphasized the role of
electronic
correlations.\cite{mottprb,paquet,ricecomment,georges,laad}

It appears that the interplay among various electronic degrees of
freedom (charge, orbital and spin) and their coupling to the
lattice degrees of freedom (phonons) leads to complex behavior in
VO$_2$. This is a hallmark of many transition-metal oxides that
exhibit remarkable properties like high-temperature
superconductivity, colossal magnetoresistance and
multiferroicity.~\cite{dagotto} Many of these transition metal
oxides have a rich phase diagram with a diversity of
phases.~\cite{dagotto} Some of the exotic properties may well be
due to phase coexistence close to phase boundaries. Recent
experimental work on high-temperature superconductors and
manganites provides direct evidence of phase separation and
inhomogeneity in these materials on mesoscopic and nanoscopic
length scales.\cite{emery,cheong,delozanne,davis,dessau,yazdani}
It is thought that competing electronic and phononic degrees of
freedom lead to phase separation and a multiplicity of phases in
transition metal oxides.~\cite{dagotto} The evolution of one phase
to another appears to proceed on the nanometer scale leading to
phase coexistence and inhomogeneity. This motivated us to
investigate the IMT regime in VO$_2$ on the nanometer length
scales with the pioneering technique of scattering scanning
near-field infrared microscopy (s-SNIM).\cite{keilmann2}

While the IMT in VO$_2$ has been studied previously by far-field
infrared and THz techniques~\cite{noh,jepsen}, our studies combine
far-field infrared spectroscopy with near-field infrared
imaging.\cite{mmqscience} With s-SNIM, we have imaged the
evolution of the thermally-induced IMT in a thin film of VO$_2$.
We found that the VO$_2$ film underwent a percolative transition
with increasing temperature across the IMT. We analyzed the s-SNIM
data along with the far-field infrared spectroscopy data within
the Bruggeman effective medium theory (EMT), and discovered that
the metallic islands embedded in the insulating host exhibit
divergent effective mass which is evidence of the Mott transition
within the Brinkman-Rice
framework.\cite{mmqscience,brinkmanrice,kimbrinkmanrice} In this
paper, we further discuss the underlying physics of the IMT in
VO$_2$. We present extensive s-SNIM images obtained while heating
and cooling across the IMT which confirm the reproducibility of
our results. We demonstrate the static nature of the coexisting
insulating and metallic regions over many minutes at a fixed
temperature. A quantitative discussion of the observed near-field
infrared amplitude contrast in the IMT regime is included. The
amplitude contrast was modeled with the so-called dipole
description~\cite{keilmann2} and its refined version. Moreover, we
explore the novel approach to data analysis of the far-field and
near-field infrared data within the Bruggeman effective medium
theory and expound upon the merits and limitations of this
procedure.

This article is divided into several sections. The details of the
sample preparation and characterization are given in the next
section. The far-field infrared spectroscopy experiments and data
analysis are described in Section \textrm{III}. This is followed
by a description of the s-SNIM technique and presentation of the
data in Sections \textrm{IV} and \textrm{V} respectively. The data
analysis within the Bruggeman effective medium theory is described
in Section \textrm{VI}. This is followed by a Summary and Outlook
Section and three appendices.

\section{Samples}

VO$_2$ films about 100 nm thick were grown on ($\bar{1}$012)
oriented sapphire (Al$_2$O$_3$) substrates by the sol gel method.
The details of growth and characterization are given in
Ref.~\onlinecite{chae}. In Fig.~\ref{sigma1}a we present the X-ray
diffraction data on the VO$_2$ film which shows that the film is
single phase $M$1-VO$_2$ with (200) orientation. The resistance of
the film was measured by a standard four-probe method, and the
data is displayed in Fig.~\ref{sigma1}b. The film undergoes a
first-order, hysteretic IMT with four orders of magnitude change
in the resistance. We chose thin films of VO$_2$ for infrared
measurements in the phase transition regime instead of single
crystals because the IMT is destructive for VO$_2$ crystals
~\cite{barkerprl} leading to potential problems of reproducibility
and significant uncertainties in the quantitative results in the
phase transition regime. Data obtained for high quality films
grown on lattice matched substrates are free of the above
complications and thus enable studies of the intrinsic properties
of VO$_2$. Our VO$_2$ films do not show signs of deterioration
even after going through several cycles across the IMT. This was
verified by repeated resistance measurements of the IMT, and by
ellipsometric and reflectance experiments that give reproducible
data in the insulating and metallic phases before and after
cycling through the IMT.

\section{Far-field infrared spectroscopy}

Spectroscopic ellipsometry provides the ellipsometric coefficients
($\Psi$($\omega$) and $\Delta$($\omega$)) at each measured
frequency.\cite{tompkins} This enables us to obtain the optical
constants of the VO$_2$ film without recourse to Kramers-Kronig
analysis which may not be applicable in the IMT regime of VO$_2$
because of inhomogeneity and phase-coexistence. Provided the
incident wavelength is large compared to the size of the
inhomogeneities, the inhomogeneous system can be described by an
effective dielectric function\cite{carrtanner}. This is certainly
expected to be the case for long wavelengths in the far- and
mid-infrared regime (between 40 cm$^{-1}$ and 5000 cm$^{-1}$).
Furthermore, we find that in accord with the EMT
\cite{carrtanner}, weighted averaging of the dielectric functions
of the insulating phase and rutile metallic phase at shorter
wavelengths (higher frequencies) adequately describes the data
(see Appendix A). Therefore, it is appropriate to assign an
effective dielectric function to the inhomogeneous,
phase-separated regime of VO$_2$ even when the wavelength becomes
comparable to the size of the metallic puddles. The effective
dielectric function  of the inhomogeneous, phase-separated regime
of VO$_2$ is thus obtained via the standard analysis of the
combined ellipsometric and reflectance data based on a two-layer
model of a VO$_2$ film on a sapphire
substrate.\cite{VO2mmq2,tompkins} The ellipsometric data spanned
the frequency range 400 - 20000 cm$^{-1}$, and near-normal
incidence reflectance covered the frequency range 40 - 680
cm$^{-1}$.

We note that in general, the complex dielectric function
$\tilde{\epsilon}$($\omega$)= $\epsilon_1$($\omega$) +
$i\epsilon_2$($\omega$) of a material is related to the complex
conductivity $\tilde{\sigma}$($\omega$)= $\sigma_1$($\omega$) +
$i\sigma_2$($\omega$) by the following equation:

\begin{equation}
\tilde{\epsilon}(\omega)=1+\frac{4\pi{i}}{\omega}\tilde{\sigma}(\omega)
\label{epsilon-sigma}
\end{equation}

The real part of the optical conductivity $\sigma_1$($\omega$) is
plotted in Fig.~\ref{sigma1}c as a function of frequency
($\omega$) for representative temperatures in the IMT regime of
VO$_2$. The inset in Fig.~\ref{sigma1}c shows the
temperature-dependence of the real part of the dielectric function
$\epsilon_1$($\omega$ = 50 cm$^{-1}$) in the IMT regime.
Fig.~\ref{sigma1}d and its inset show $\sigma_1$($\omega$) and
$\epsilon_1$($\omega$ = 50 cm$^{-1}$) respectively with phonon
contributions subtracted. The contributions of VO$_2$ phonons to
the dielectric functions were modeled by Lorentzian oscillators
and subsequently subtracted so that the electronic contribution to
the optical constants could be unambiguously presented and
analyzed.

An optical gap of $\approx$ 4000 cm$^{-1}$ (0.5 eV) is evident in
$\sigma_1$($\omega$) of the insulating phase (295 K $\leq T \leq$
341 K). As the VO$_2$ film enters the IMT regime for $T >$ 341 K,
the conductivity in the gap increases and gradually fills the gap.
The first inter-band transition centered at 10500 cm$^{-1}$ does
not appear to shift to lower frequencies as would be expected if
the gap were to collapse. The filling up of the optical gap,
rather than a collapse, is also seen in other correlated (or Mott)
insulators like the cuprates that are rendered conducting through
chemical doping.\cite{imada-rmp,basovreview,sashareview} In
addition, we observe an isosbestic point in $\sigma_1$($\omega$)
characterized by invariant conductivity (depicted by an open
circle in Fig.~\ref{sigma1}c,d). The isosbestic point is also seen
in other Mott systems~\cite{tokura} although its significance is
yet to be understood. The observations discussed in the preceding
sentences point to the importance of electronic correlations in
VO$_2$. As the temperature is increased, the low frequency
conductivity increases until the film becomes fully metallic as
seen from the broad Drude-like feature at $T$ = 360 K. We note
that the enhanced spectral weight at low energies as the film
becomes increasingly metallic is borrowed from higher lying
inter-band transitions.\cite{okazaki,VO2mmq1,VO2mmq2} The
divergence of $\epsilon_1$ shown in the insets of
Fig.~\ref{sigma1}c,d is similar to that known from percolative
insulator-to-metal transitions where it occurs due to enhanced
capacitive coupling between the metallic regions as they grow and
proliferate in the insulating host.\cite{noh,jjtu} $\epsilon_1$
abruptly becomes negative at $T$ = 343 K signaling that the
metallic regions have percolated (connected) forming a macroscopic
conducting path for the charge carriers.

\section{Scanning near-field infrared microscopy}

The characteristic evolution of the effective area-averaged
optical constants in the IMT regime of VO$_2$ provides indirect
evidence of a percolative phase transition with coexisting
insulating and metallic regions. Our objective was to directly
observe the percolation with scanning near-field optical
techniques. We chose to directly image the IMT with s-SNIM at the
mid-infrared frequency of 930 cm$^{-1}$. As can be seen from
Fig.~\ref{sigma1}c,d, the optical conductivity of the insulating
and metallic phases of VO$_2$ is very different at this frequency
and so a large contrast is expected in the near-field amplitude
signal.

Scattering scanning near-field infrared microscopy was performed
with a custom Atomic Force Microscope (AFM) using Pt-coated
silicon tips in the tapping mode.\cite{keilmann2,keilmann} The
$\omega$ = 930 cm$^{-1}$ infrared frequency was provided by a
CO$_2$ laser. The pseudo-heterodyne scheme with second harmonic
demodulation was employed to isolate and detect the near-field
amplitude signal.\cite{hillenbrand} The spatial resolution of the
near-field infrared probe is typically 10-20 nm and is set by the
radius of curvature of the tip. The near-field penetrates about 20
nm into the VO$_2$ film and can therefore be considered a bulk
probe of the insulator-to-metal transition.\cite{taubner}

The near-field interaction between the tip and the sample is
described by the dipole model. This is the simplest model that
quantitatively describes the near-field interaction and has been
found to be accurate in some cases.~\cite{keilmann2} Within this
model, the complex scattered signal at the second harmonic of the
tapping frequency $\Omega$ is a function of the optical constants
of the tip and the sample.~\cite{keilmann2} The incident infrared
light polarizes the tip and within the dipole model, it is
sufficient to model the polarizability of the tip with an
effective spherical dipole of radius $a$ ($\approx$ 20 nm). The
interaction between the tip dipole and the sample is modeled by
considering a mirror dipole in the sample whose polarizability
depends on the local optical constants of the sample in the region
directly below the tip. The effective polarizability of the
interacting tip-sample system ($\alpha_{eff}$) is given by:

\begin{equation}
\alpha_{eff}=\frac{\alpha(1+\beta)}{1-\frac{\alpha\beta}{16\pi(a+z(t))^3}}.
\label{dipole}
\end{equation}

$\alpha_{eff}$ is complex-valued and relates the incident $E_i$
and scattered $E_s$ electric fields via $E_s \propto
\alpha_{eff}E_i$. The polarizability of the tip $\alpha$ is given
by $\alpha =
4\pi{a^3}(\tilde{\epsilon}_t-1)/(\tilde{\epsilon}_t+2)$ where
$\tilde{\epsilon}_t$ is the complex dielectric function of the tip
at the incident laser frequency. The polarizability of the mirror
dipole in the sample is given by $\alpha\beta$ where $\beta =
(\tilde{\epsilon}_s-1)/(\tilde{\epsilon}_s+1)$ expresses the local
dielectric response of the sample that depends on its complex
dielectric function $\tilde{\epsilon}_s$ at the incident laser
frequency.

The effective polarizability of the tip-sample system is a
non-linear function of the tip-sample distance $z$. The tip-sample
distance varies sinusoidally with time $z(t) =
z_0(1+\cos(\Omega{t}))$ due to oscillation of the tip. In the
tapping mode, the tip oscillates with an amplitude $z_0$ of about
40 nm at a frequency $\Omega$ = 25 kHz. Therefore, the effective
polarizability and hence the near-field signal is modulated at
harmonics of $\Omega$.

\begin{align}
\alpha_{eff}=(\alpha_{eff})_0+(\alpha_{eff})_1\cos(\Omega{t})+(\alpha_{eff})_2\cos(2\Omega{t})\nonumber\\
+(\alpha_{eff})_3\cos(3\Omega{t})+...+
(\alpha_{eff})_n\cos(n\Omega{t}). \label{expansion}
\end{align}

The $n$th harmonic coefficient of the near-field polarizability is
as follows:

\begin{equation}
(\alpha_{eff})_n=\frac{\Omega}{2\pi}\int_{-\pi/\Omega}^{\pi/\Omega}{\alpha_{eff}\cos{(n\Omega{t})}dt}.
\label{fourier}
\end{equation}

Here $\alpha_{eff}$ is given by eq.~\ref{dipole}. The $n$th
harmonic of the scattered electric field $(E_s)_n = s_ne^{i\phi_n}
\propto (\alpha_{eff})_nE_i$, where $s_n$ and $\phi_n$ are the
$n$th harmonic amplitude and phase coefficients respectively. Thus
the second harmonic coefficient of the near-field amplitude $s_2$
can be calculated (up to a proportionality constant that can be
calibrated) provided the complex dielectric functions of the tip
and sample at the incident laser frequency are known.

More recently, it has been suggested that the (1+$\beta$) term in
the numerator of eq.~\ref{dipole} be replaced by the term (1 +
$r$)$^2$, where $r$ is the far-field Fresnel reflectance
coefficient:\cite{raschke,markusthesis,ocelic}

\begin{equation}
\alpha_{eff}=\frac{\alpha(1+r)^2}{1-\frac{\alpha\beta}{16\pi(a+z(t))^3}}.
\label{dipole2}
\end{equation}

In the above equation, the (1 + $r$)$^2$ term accounts for the
indirect illumination of the tip from reflections at the sample
surface.

The dielectric function $\tilde{\epsilon}_t$ of Pt at $\omega$ =
930 cm$^{-1}$ is $-1300+960i$.\cite{ordal} The dielectric
functions of VO$_2$ at $\omega$ = 930 cm$^{-1}$ are obtained from
ellipsometric chracterization. These are as follows: 4.9 with
negligibly small imaginary part for the monoclinic insulating
phase, and $-35+119i$ for the rutile metallic phase. Therefore,
the scattered near-field amplitude in the metallic phase of VO$_2$
is expected to be about twice that in the insulating phase from
basic dipole model (eq.~\ref{dipole}). However, the extended
dipole model (eq.~\ref{dipole2}) gives a factor of 3 ratio for the
near-field amplitudes in the metallic and insulating phases.

\section{Infrared nano-imaging of the insulator-to-metal transition
 in VO$_2$}

We now turn to Fig.~\ref{images} which displays the second
harmonic amplitude images in the IMT regime of VO$_2$ obtained
with s-SNIM at the incident laser frequency of 930 cm$^{-1}$. At
$T$ = 341 K, we observe an essentially uniform amplitude signal.
As the temperature is increased, nanometer-scale regions with
higher near-field amplitude appear. The increased near-field
amplitude of these nano-scale puddles is due to their metallicity.
The metallic puddles subsequently proliferate and grow in size,
and then coalesce with increasing temperature. The $T$ = 360 K
near-field map shows higher near-field amplitude indicating that
the scanned region has become completely metallic. Near-field
amplitude images were also obtained while cooling the VO$_2$ film
through the IMT and are shown in Appendix B. We note here that
coexisting insulating and metallic domains in VO$_2$ nano-rods in
the IMT regime have been observed with optical microscopy in
Ref.~\onlinecite{park}. Moreover, the IMT at the surface of VO$_2$
films has also been imaged with scanning tunneling
microscopy.~\cite{nohstm}

The distribution of the scattered near-field amplitude  for
various temperatures in the IMT regime of VO$_2$ is plotted in
Fig.~\ref{histograms}a. This distribution is obtained from the
images in Fig.~\ref{images}. The asymmetry of the histograms and
the presence of two peak structures are evidence of coexisting
insulating and metallic phases in the IMT regime of VO$_2$. The
peak at lower near-field amplitudes corresponds to the insulating
phase and the one at higher amplitudes corresponds to the metallic
phase. The median amplitude for the metallic phase at $T$ = 360 K
is about 15 times higher than the median amplitude for the
insulating phase. This is far higher than the factor of 2-3
expected from the dipole model. While eq.~\ref{dipole2} improves
the description of the near-field contrast in our data compared to
eq.~\ref{dipole}, it still underestimates the contrast compared to
experiment. A similar discrepancy between the measured near-field
contrast and that predicted by the dipole model has also been
noted for near-field terahertz measurements on the IMT in
VO$_2$.\cite{mittleman}

We note that each of separate images in Fig.~\ref{images} has only
a factor of 2-3 variation of the near-field amplitude contrast.
Moreover, the median position of the ``insulating" peak in
Fig.~\ref{histograms}a shows a small but noticeable increase with
temperature without apparent reason. These observations led us to
normalize the measured amplitude to the peak position of the
insulating phase. The resulting distributions plotted as a
function of the normalized amplitude are in fairly good agreement
with the predictions of the dipole models (see
Fig.~\ref{histograms}b). The amplitude from the metallic phase is
now a factor of 2-4 higher than that from the insulating phase.
Note that the basic dipole model (eq.~\ref{dipole}) gives better
quantitative agreement in the regime with smaller sizes of the
metallic puddles in the insulating host and vice versa. The
extended dipole model (eq.~\ref{dipole2}) explains the data well
in the regime where the insulating and metallic regions have
similar length scales. The apparent continuous increase in the
near-field amplitude of the metallic phase with increasing
temperature (see Fig.~\ref{histograms}a) and the 15 times higher
near-field amplitude from the metallic phase compared to the
insulating phase are phenomena that are not fully understood at
present and require further investigation. The insulating and
metallic fractions can, nevertheless be extracted from the
near-field images and amplitude distributions as explained in the
next section.

We note that the near-field images are repeatable when taken in
five minute sequences over the same sample area and at a fixed
temperature because they show nearly identical patterns of
metallic puddles in the insulating host (see Appendix B). This
indicates that the possible effect of dynamic fluctuations on the
static patterns is small. This is discussed in detail in Appendix
B.

\section{Broad-band infrared response of the metallic regions in the phase coexistence regime of VO$_2$}

As shown in Appendix A, weighted averaging of optical constants of
the insulating phase and of the rutile metallic phase within the
Bruggeman effective medium theory (EMT) produces a good
description of the far-field infrared data near the onset of the
insulator-to-metal transition at high frequencies. Deviations from
this description become progressively stronger at longer
wavelengths, a regime where the EMT formalism is expected to be
best applicable.\cite{carrtanner} This implies that the optical
constants of the metallic puddles when they first appear at the
onset of the insulator-to-metal transition are different from the
optical constants of the rutile metal at $T$ = 360 K. Therefore,
we use the EMT to extract the optical constants of the metallic
puddles.

Within the Bruggeman effective medium theory,\cite{carrtanner} the
effective (complex) optical constants
$\tilde{\epsilon_E}$($\omega$) of a two-component inhomogeneous
system are given by:

\begin{equation}
f\frac{\tilde{\epsilon_a}(\omega)-\tilde{\epsilon_E}(\omega)}
{\tilde{\epsilon_a}(\omega)+\frac{1-q}{q}\tilde{\epsilon_E}(\omega)}
+(1-f)\frac{\tilde{\epsilon_b}(\omega)-\tilde{\epsilon_E}(\omega)}
{\tilde{\epsilon_b}(\omega)+\frac{1-q}{q}\tilde{\epsilon_E}(\omega)}
= 0 \label{EMTequation}
\end{equation}

In this equation, $\tilde{\epsilon_a}$($\omega$) and
$\tilde{\epsilon_b}$($\omega$) are the complex optical constants
of the metallic and insulating phases respectively, \emph{f} and
(1-\emph{f}) are the volume fractions of the metallic and
insulating phases respectively, and \emph{q} is the depolarization
factor that depends on the shape of the components which is
inferred from near-field images. This factor is taken to be
0.2-0.4 assuming nearly spherical metallic regions at low
concentrations of the newborn metallic state. This is because
their lateral dimensions are less than the film thickness and the
out-of-plane dimension is assumed to be similar to the lateral
dimensions. The depolarization factor continuously increases to
0.5 (appropriate for thin flat disks) at higher concentrations of
metallic clusters as their lateral dimensions exceed the
out-of-plane dimension which is limited by the thickness of the
film. The volume fractions of the insulating and metallic phases
are obtained directly from the near-field images by fitting the
amplitude distribution curves in Fig.~\ref{histograms}a to two
Gaussians, with the area under each Gaussian yielding the volume
fraction of the respective phase. The optical constants of the
insulating phase $\tilde{\epsilon_b}$($\omega$) and the
inhomogeneous transition regime $\tilde{\epsilon_E}$($\omega$)
have also been measured (see Fig. 1). The optical constants of the
insulating phase are assumed to be independent of temperature in
this analysis. Since $\tilde{\epsilon_E}$($\omega$),
$\tilde{\epsilon_b}$($\omega$), and \emph{f} are known whereas the
range of \emph{q} is constrained based on the knowledge of the
metallic cluster size and film thickness, the equation can be
solved to obtain the optical constants of the metallic regions
$\tilde{\epsilon_a}$($\omega$), and hence the complex conductivity
$\tilde{\sigma_a}$($\omega$) for each temperature. The real part
of the conductivity $\sigma_{1a}$($\omega$) extracted by this
procedure is plotted in Fig.~\ref{sigmametal}(a) for
representative temperatures in the IMT regime. The values of $f$
and $q$ used for the analysis are listed in Table~\ref{tfq}.

\begin{table}
\begin{tabular*}{60mm}{c@{\extracolsep{15mm}}
c@{\extracolsep{15mm}}c}
\hline \\
$T$ (K) & $f$ & $q$ \\
\hline \\
342 & 0.18 & 0.2 \\
342.6 & 0.31 & 0.33 \\
343 & 0.48 & 0.45 \\
343.6 & 0.7 & 0.5 \\
\hline \\
\end{tabular*}
\caption{The values of the filling fraction $f$ and depolarization factor
$q$ used to obtain the optical constants of the metallic regions at selected
temperatures within the framework of the EMT formalism.}
\label{tfq}
\end{table}

We note that the form of $\sigma_{1a}$($\omega$) for the metallic
regions that are separated by the insulating host exhibits a
narrow Drude-like peak at low frequencies followed by a
``dip-hump" structure at higher frequencies. This is quite
different from $\sigma_{1a}$($\omega$) of the rutile metal at $T$
= 360 K which has a broad and featureless Drude peak. As the
temperature is increased, and the metallic regions proliferate and
grow, their $\sigma_{1a}$($\omega$) rapidly evolves to that seen
in the rutile metallic phase. Here we note that the extracted
optical constants of the metallic puddles at $\omega$ = 930
cm$^{-1}$ are different from the optical constants of the $T$ =
360 K rutile metal at this frequency. Nevertheless, within the
dipole model (eqs.\ref{dipole} and \ref{dipole2}), the near-field
scattering amplitude at $\omega$ = 930 cm$^{-1}$ for the metallic
puddles is expected to be nearly the same as that from the $T$ =
360 K rutile metal.

Uncertainties in the magnitude of \emph{q} do affect the behavior
of the extracted spectra of the metallic puddles at the lowest
frequencies accessible in our experiment. This is shown in
Fig.~\ref{sigma1puddles}. Changing \emph{q} for a fixed value of
\emph{f} results in a variation in the $\sigma_1$($\omega$)
spectra below 800 cm$^{-1}$ whereas the variation is negligibly
small above 800 cm$^{-1}$. We see that the mid-infrared band
(``hump") peaked at $\approx$ 1800 cm$^{-1}$ and the ``dip"
feature below this frequency are robust features independent of
the value of \emph{q}. Below 300 cm$^{-1}$ however, the choice of
$q$ produces either a Drude-like feature or a finite energy mode
which is consistent with Drude dynamics modified by
localization.\cite{basovreview} The latter behavior is often seen
in systems close to the metal-insulator transition in the presence
of disorder.\cite{basovprl98,dummprl2002,dummprl2003} We note that
for a value of \emph{f} obtained from the near-field image at a
fixed temperature, the range of \emph{q} values is highly
constrained. Within EMT, percolation is predicted for \emph{f}
$\geq$ \emph{q}.\cite{carrtanner} However, the far-field effective
spectra and the near-field patterns for $T$ = 342.6 K indicate
that percolation has not yet occurred and this means that \emph{f}
$<$ \emph{q}. Since solutions of eq.\ref{EMTequation} yielding a
finite energy mode occur for \emph{f} $\geq$ \emph{q}, we infer
that these are unlikely to be correct, although they cannot be
dismissed entirely. We also see that for \emph{q} $>$ 0.34,
$\sigma_{1a}$ turns negative over a finite frequency range and is
unphysical. Hence, the spectra showing a narrow Drude-like mode
centered at zero frequency are the most probable candidates for
the optical constants of the metallic puddles.

The scattering rate 1/$\tau$($\omega$) and the mass enhancement
factor $m$*($\omega$)/$m_b$ of the metallic puddles obtained from
the extended Drude analysis of the optical constants (see Appendix
C) are plotted in Fig.~\ref{sigmametal}(b),(c) for representative
temperatures in the IMT regime. The scattering rate exhibits a
gap-like form at low frequencies with an overshoot at $\approx$
1000 cm$^{-1}$. We attribute the suppression of the scattering
rate at low frequencies to partial gapping of the Fermi surface
(i.e. a pseudogap). The connection between the suppression in the
scattering rate and the opening of a pseudogap has been made
earlier for the under-doped cuprates that are close to a Mott
transition.\cite{basovprl96,puchkovprl96,singley} The pseudogap in
VO$_2$ could be due to Mott physics, the Peierls instability
\cite{gruner} or a complex interplay between the two. We cannot
rule out the possibility that this peculiar form of
1/$\tau$($\omega$) and the ``dip-hump" structure in
$\sigma_1$($\omega$) are due to a low-lying interband transition.

Note that the mass enhancement factor of the metallic puddles
increases substantially at low frequencies. The mass enhancement
factor in the limit of $\omega$ $\rightarrow$ 0 diverges as one
approaches the IMT with varying temperature (see inset of
Fig.~\ref{sigmametal}(c)). The mass divergence is likely due to
electron-electron interactions and is a fingerprint of a Mott
transition.\cite{brinkmanrice,kimbrinkmanrice} Note that our
observation of mass divergence is not inconsistent with the
first-order nature of the phase transition in VO$_2$ because about
two-thirds of the entropy associated with the first-order
transition is related to the lattice and the rest is
electronic.~\cite{mottprb} The relationship of electronic mass
divergence and the electronic IMT to the structural transition
needs to be explored in the future with nanoscale X-ray
diffraction measurements of the metallic puddles.

As mentioned in the preceding paragraph, the form of the
conductivity below 300 cm$^{-1}$ is somewhat ambiguous. While the
narrow Drude-like solutions show a variation in the low frequency
conductivity, all such solutions lead to enhanced effective mass
and an optical pseudogap in the metallic puddles within the
extended Drude analysis. For solutions exhibiting a low energy
mode, the reduced oscillator strength of this low-energy mode
compared to that of the Drude mode in the rutile metal indicates
enhanced optical effective mass.

Interfacial scattering and localization effects will become
important when the size of the metallic puddles is comparable to
or less than the intrinsic mean free path of the charge carriers
within these puddles. Our estimate of the mean free path based on
the scattering rate data in Fig.~\ref{sigmametal}(b) is of the
order of a few nanometers. The smallest puddles we can detect
(limited by the 20 nm spatial resolution of our probe) are still
in the regime where the mean free path is shorter than the spatial
extent of metallic regions. Therefore, interfacial scattering and
localization are unlikely to affect charge dynamics in the
metallic puddles. However, we cannot rule out the possibility of
quasiparticle localization which normally leads to a finite energy
mode in the optical conductivity spectrum.

\section{Summary and outlook}

We have performed far-field infrared ellipsometry and reflectance
measurements on VO$_2$ films across the insulator-to-metal
transition (IMT) and have determined the evolution of the
effective optical constants. In addition, we have investigated the
IMT with scanning near-field infrared microscopy and have directly
observed the percolative phase transition. A combination of the
far-field and near-field results within the ambit of the Bruggeman
effective medium theory uncovers enhanced effective mass in the
incipient metallic islands at the onset of the IMT. This result
signifies the pivotal role of electronic correlations in the IMT
physics.

We find that the dipole model of the near-field infrared contrast
provides a satisfactory explanation of the observed amplitude
contrast between the metallic puddles and the insulating regions.
The agreement between experiment and the dipole model is evident
only after the near-field amplitude from the metallic regions is
normalized to the near-field amplitude from the insulating
regions. Future s-SNIM experiments on VO$_2$ with Au reference
should help better understand the evolution of the near-field
amplitude as the sample goes through the IMT.

The methods and results reported here set a new standard for
investigating the conducting state in other doped correlated
insulators, for example the cuprates and manganites, where phase
segregation is
commonplace.\cite{emery,cheong,delozanne,davis,dessau,yazdani}
Ideally, one wants to obtain broad-band near-field infrared
characterization on the nano-scale, without recourse to effective
medium theory. The technology for this purpose is in its infancy
and requires development.\cite{broadband,huber,naturenews} Our
work provides an incentive for such endeavors.

\section{Acknowledgements}

This work was supported by the U.S. Department of Energy Grant
No.DE-FG03-00ER45799, the Deutsche Forschungsgemeinschaft Cluster
of Excellence Munich – Centre for Advanced Photonics, and the
Electronics and Telecommunications Research Institute (ETRI),
Korea.

\section{Appendix A}

The measured ellipsometric coefficients $\Psi$($\omega$) and
$\Delta$($\omega$) for $T$ = 342 K are compared in
Fig.~\ref{psidelta}a,b with the results from Bruggeman effective
medium theory for an inhomogeneous two-phase system.  The latter
is a fit to the experimental data based on the optical constants
of the monoclinic ($M1$) insulator and the rutile metal at $T$ =
360 K. The filling fractions and depolarization factor used for
the fits are taken from Table~\ref{tfq}. Satisfactory agreement is
obtained at higher frequencies (10000 - 20000 cm$^{-1}$) as can be
seen clearly in the plots of the difference between the fits and
the data displayed in the insets of Fig.~\ref{psidelta}a,b.
However, the Bruggeman EMT should produce the most accurate
results at low frequencies which is not the case here. This
implies that the optical constants of the metallic phase in the
inhomogeneous IMT regime may be different from those of the rutile
metal at $T$ = 360 K provided the optical constants of the $M$1
insulator do not change in the IMT regime.

\section{Appendix B}

Near-field amplitude images obtained while cooling the VO$_2$ film
through the IMT are shown in Fig.~\ref{cooling}. The images show
that the insulating phase nucleates in the metallic host, and then
proliferates and finally percolates. The IMT occurs about 8 K
lower in temperature than on heating. Such a hysteresis is
expected for a first-order phase transition and is seen in
Fig.\ref{sigma1}b.

In Fig.~\ref{static}a,b we demonstrate that repeated near-field
imaging at fixed conditions yields unchanged patterns of metallic
puddles. Imaging takes about 5 minutes ($\approx$ 0.02 s per pixel
of $\approx$ 15 nm $\times$ 15 nm size) and was immediately
repeated. The reproducible patterns of metallic puddles rules out
the possibility of slow dynamic fluctuations. If dynamic
fluctuations occur on shorter time scales, then they do not affect
the average pattern, especially they do not nucleate and stabilize
into static metallic puddles.\cite{LandauLifshitz} We hypothesize
that the specific patterns that emerge are most likely seeded and
controlled by strain at the interface, defects, grain boundaries
etc. These nucleation centers should lead to a reproducible
pattern at a particular temperature, a feature that could not
however be proven in the experimental setup because thermal drift
due to thermal cycling through the phase transition prevented the
repeated s-SNIM imaging of the same sample area.

We note here that simultaneous topography images were obtained
along with near-field infrared images displayed in this work. We
verified that the near-field infrared contrast between insulating
and metallic regions in the IMT regime is due to the difference in
optical constants and not due to topographic artifacts or changes
in topography. However, we were unable to study our hypothesis
that the metallic islands are seeded at grain boundaries, defects
etc. This is because extensive image processing is required in
order to identify the relationship between the new-born metallic
islands and grain boundaries. This analysis can be reliably
applied only if the topography data are more detailed. Optimized
topography images will be required to investigate possible
correlations between the nucleation sites of metallic islands and
topographic features like grain boundaries.

\section{Appendix C}

The following equations of the extended Drude formalism
~\cite{basovreview} describe the procedure for extracting the
scattering rate 1/$\tau$($\omega$) and the mass enhancement factor
$\emph{m}$*($\omega$)/$\emph{m}_b$ from the real and imaginary
parts of the complex optical conductivity of the metallic regions
$\tilde{\sigma_{a}}$($\omega$)= $\sigma_{1a}$($\omega$) +
$i\sigma_{2a}$($\omega$):

\begin{align}
\frac{1}{\tau(\omega)}=\frac{\omega_p^2}{4\pi}\Big(\frac{\sigma_{1a}(\omega)}{\sigma_{1a}^2(\omega)+\sigma_{2a}^2(\omega)}\Big)\\
\frac{m^*(\omega)}{m_b}=\frac{\omega_p^2}{4\pi\omega}\Big(\frac{\sigma_{2a}(\omega)}{\sigma_{1a}^2(\omega)+\sigma_{2a}^2(\omega)}\Big)
\label{xtendDrude}
\end{align}

The plasma frequency $\omega_{p}$ for the metallic states is
obtained from the partial sum rule:

\begin{equation}
\frac{\omega{_p}{^2}}{8}=\int_{0}^{\omega_c}\sigma_{1a}(\omega)d\omega
\label{sumrule}
\end{equation}

Here, the upper frequency limit $\omega_{c}$ = 13700 cm$^{-1}$ is
chosen for the rutile metal to exclude contributions from
higher-lying optical transitions.\cite{VO2mmq1,VO2mmq2} This gives
$\omega_{p}$ = 22000 cm$^{-1}$ for the rutile metal and this value
is nearly constant (within ten percent) for the metallic state in
the nanoscale puddles with fixed $\omega_{c}$ = 13700 cm$^{-1}$.
With this nearly constant choice of the plasma frequency for all
metallic states, the mass enhancement factor for the strongly
correlated metallic puddles can be directly compared to the mass
enhancement factor of the rutile metal.

The plasma frequency is related to the carrier density (\emph{n})
and the band mass ($m_b$) by the following expression:

\begin{equation}
\omega{_p}{^2}=\frac{4\pi{n}{e^2}}{m_b} \label{plasma}
\end{equation}

We assume that the carrier density does not change significantly
in the strongly correlated metallic puddles compared to the rutile
metal.

\begin{figure*}[t]
\caption{(color online)(a) Room temperature X-ray diffraction data
for the (200)-oriented $M$1-VO$_2$ film on ($\bar{1}$012)-oriented
sapphire substrate. (b) The resistance of the VO$_2$ film is
plotted while heating and cooling across the IMT. (c) Real part of
the conductivity $\sigma_1$($\omega$) of VO$_2$ plotted as a
function of frequency for various representative temperatures
while heating the sample through the IMT. The open circle denotes
the isosbestic (equal conductivity) point for all spectra. (Inset)
The temperature dependence of the real part of the dielectric
function $\epsilon_1$ in the low-frequency limit ($\omega$ = 50
cm$^{-1}$). (d) The optical constants of VO$_2$ as in (c) but with
phonon contributions subtracted.} \label{sigma1}
\end{figure*}

\begin{figure*}
\caption{(color online) Images of the near-field scattering
amplitude $s_2$ (scale in relative units) over the same 4 $\mu$m
by 4 $\mu$m area obtained by s-SNIM operating at the infrared
frequency $\omega$ = 930 cm$^{-1}$. The images were obtained for
various temperatures while heating VO$_2$ through the IMT. The
metallic regions depicted in light blue (light gray) and white
give higher amplitude compared to the insulating phase shown in
dark blue color (dark gray).} \label{images}
\end{figure*}

\begin{figure*}
\caption{(color online): (a) The distribution of near-field
amplitude $s_2$ from images in Fig.~\ref{images} for selected
temperatures. Note the log scale on the vertical axis for clarity
of presentation. (b) the distribution of the normalized near-field
amplitude for the same temperatures as in (a). The median
near-field amplitudes for the insulating phase were used for
normalizing the near-field amplitude in (b). The distributions
plotted in panels (a) and (b) are the number of pixels normalized
such that the area under each of the histograms is unity.}
\label{histograms}
\end{figure*}

\begin{figure*}
\caption{(color online) Optical characterization of the metallic
puddles derived from experiment: (a) the real part of the optical
conductivity $\sigma_{1a}$($\omega$) (b) the scattering rate
1/$\tau$($\omega$) and (c) the optical effective mass normalized
by the band value $m*$($\omega$)/$m_b$ of the metallic regions of
VO$_2$ for representative temperatures in the IMT regime. The
inset in (c) shows the $\omega$ $\rightarrow$ 0 limit of the mass
enhancement factor as a function of temperature (data points
between T = 400 K and 550 K are from
Ref.\onlinecite{VO2mmq1}).}\label{sigmametal}
\end{figure*}

\begin{figure*}
\caption{(color online) Dependence of the real part of the optical
conductivity $\sigma_{1a}$($\omega$) of the metallic puddles on
the depolarization factor $q$, with $f$ fixed at 0.31 for $T$ =
342.6 K.}\label{sigma1puddles}
\end{figure*}

\begin{figure*}
\caption{Panels (a) and (b) display the ellipsometric coefficients
$\Psi$($\omega$) and $\Delta$($\omega$) respectively that were
measured for the VO$_2$ film on a sapphire substrate at $T$ = 342
K (dashed gray curves). The thin black solid lines are the
ellipsometric coefficients derived from the Bruggeman effective
medium theory model as explained in Appendix A. The insets show
the difference between the fits and the data for the respective
ellipsometric coefficients.} \label{psidelta}
\end{figure*}

\begin{figure*}
\caption{(color online) Near-field amplitude $s_2$ images (scale
in relative units) obtained at $\omega$ = 930 cm$^{-1}$ with
decreasing temperature through the IMT.} \label{cooling}
\end{figure*}

\begin{figure*}
\caption{(color online) Near-field amplitude $s_2$ images at
$\omega$ = 930 cm$^{-1}$ at $T$ = 342 K. Image (b) was obtained
immediately after image (a).} \label{static}
\end{figure*}

\begin{figure*}[t]
\includegraphics[width=120mm,height=150mm,bb=72 245 550 750]{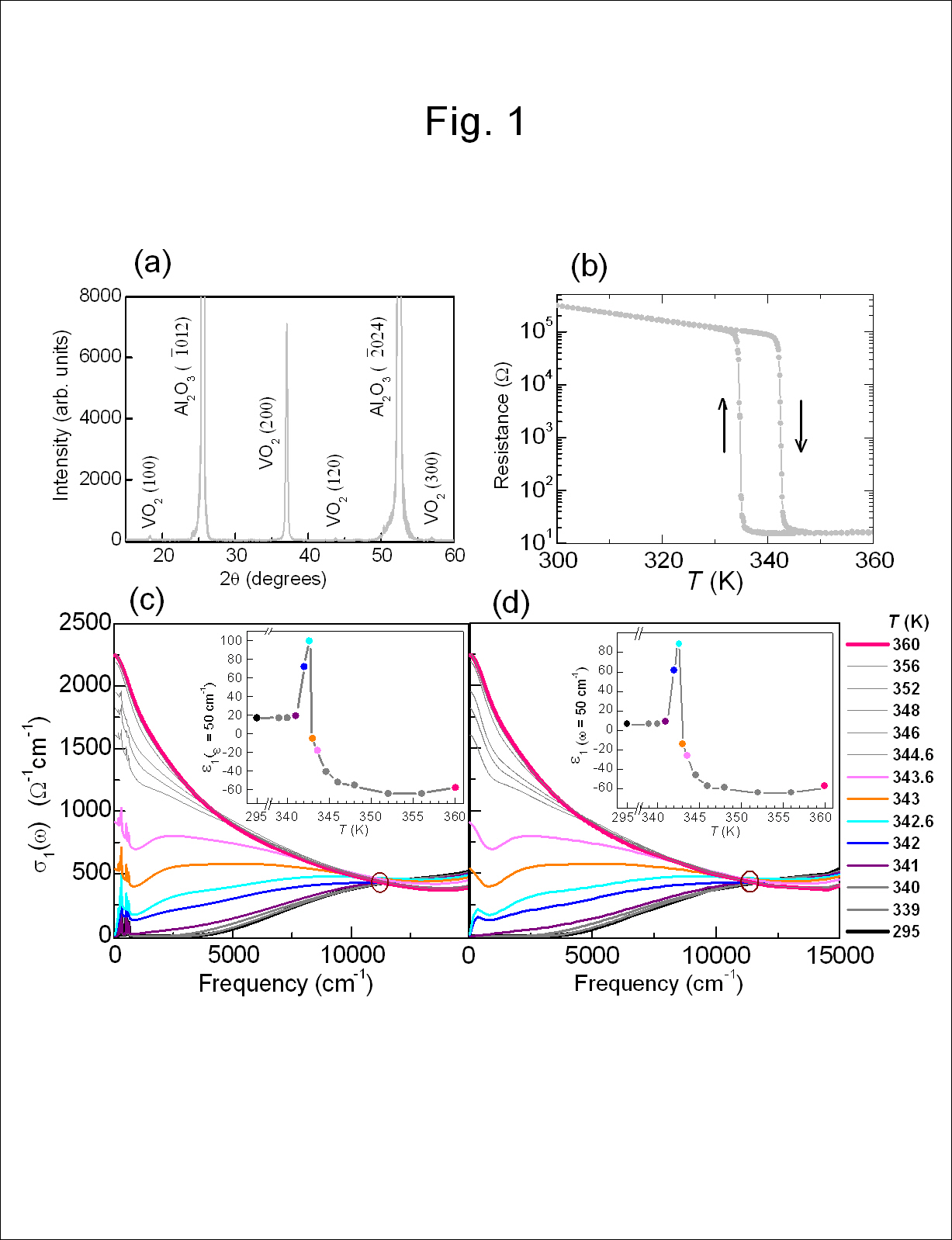}
\end{figure*}

\begin{figure*}[t]
\includegraphics[width=150mm,bb=0 144 620 750]{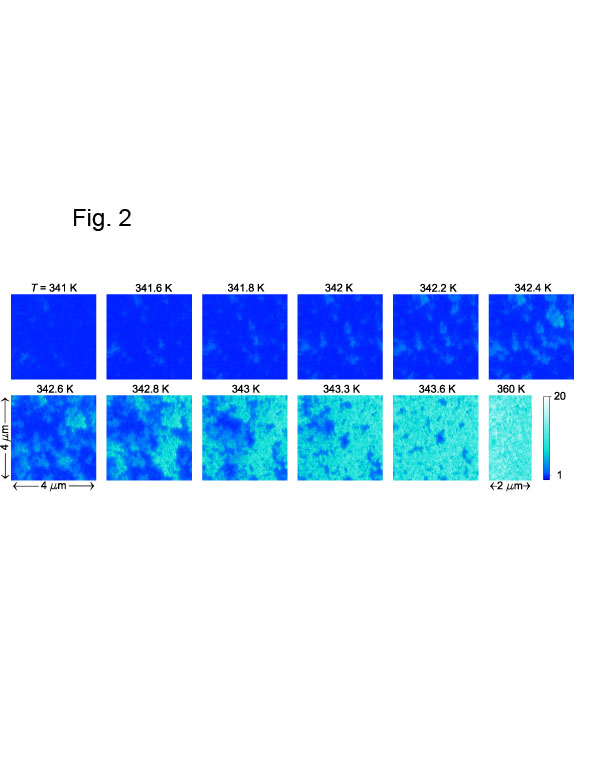}
\end{figure*}

\begin{figure*}[t]
\includegraphics[width=150mm,height=160mm,bb=72 245 550 750]{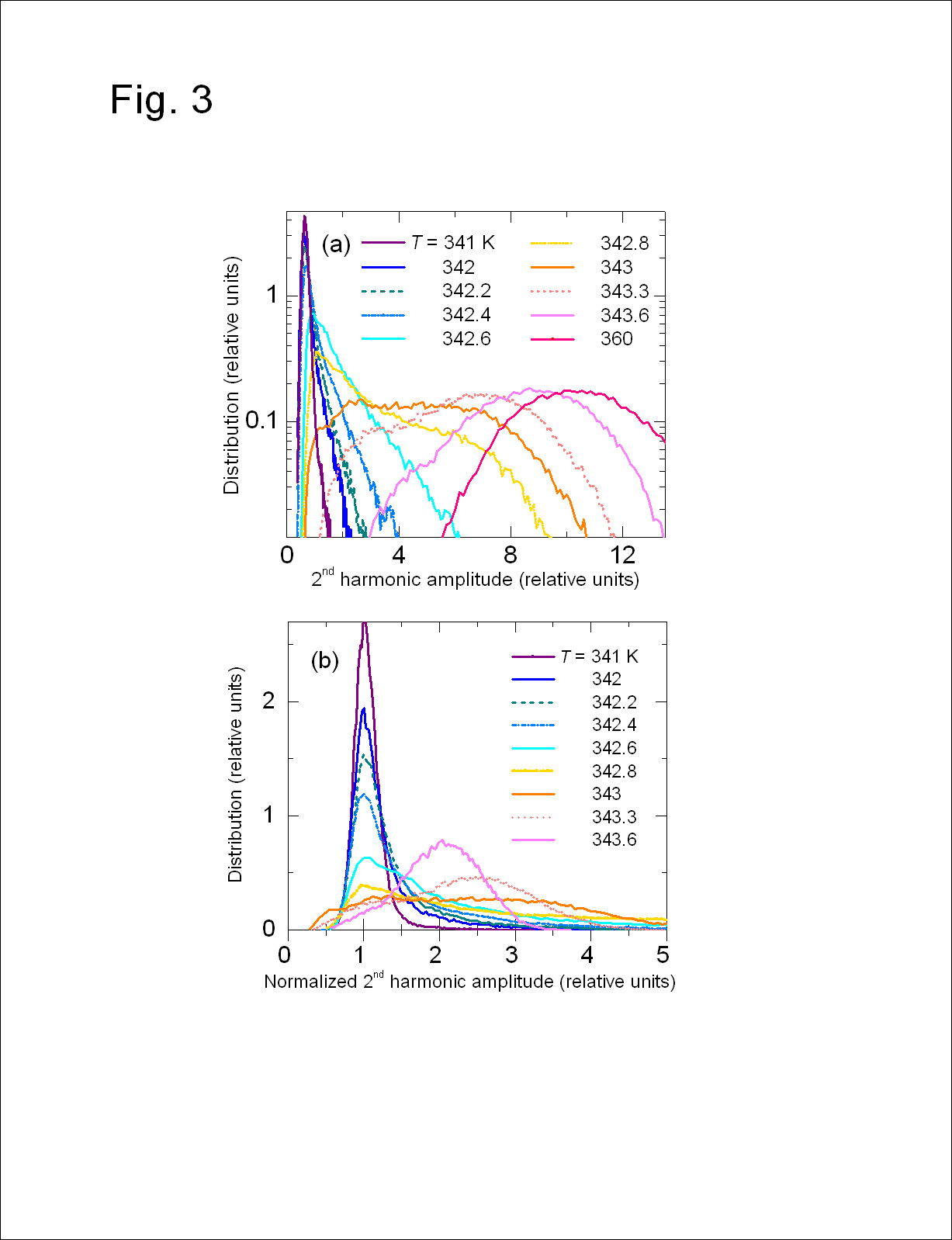}
\end{figure*}

\begin{figure*}[t]
\includegraphics[width=150mm,height=160mm,bb=72 245 550 750]{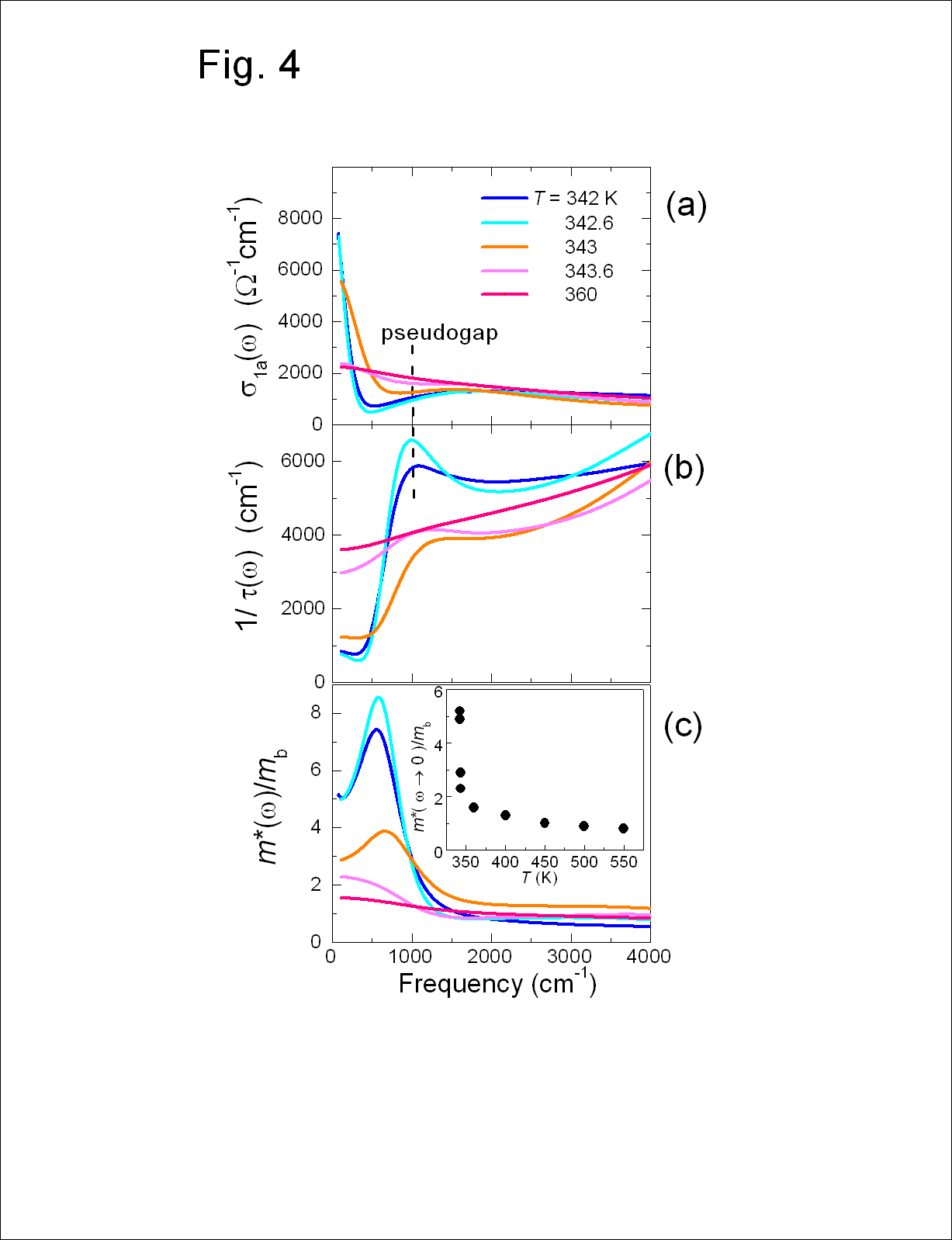}
\end{figure*}

\begin{figure*}[t]
\includegraphics[width=120mm,height=150mm,bb=72 245 550 750]{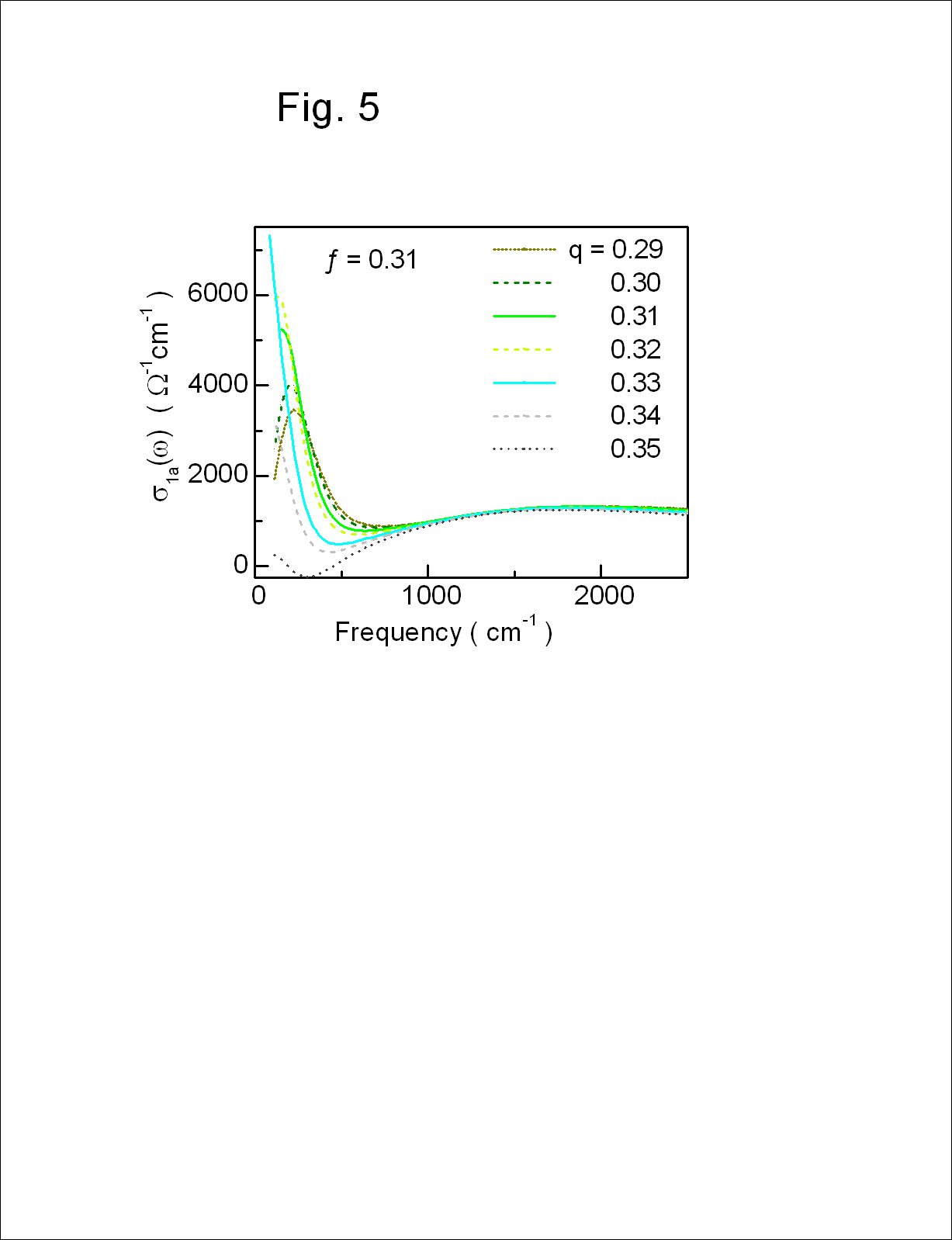}
\end{figure*}

\begin{figure*}[t]
\includegraphics[width=120mm,height=150mm,bb=72 245 550 750]{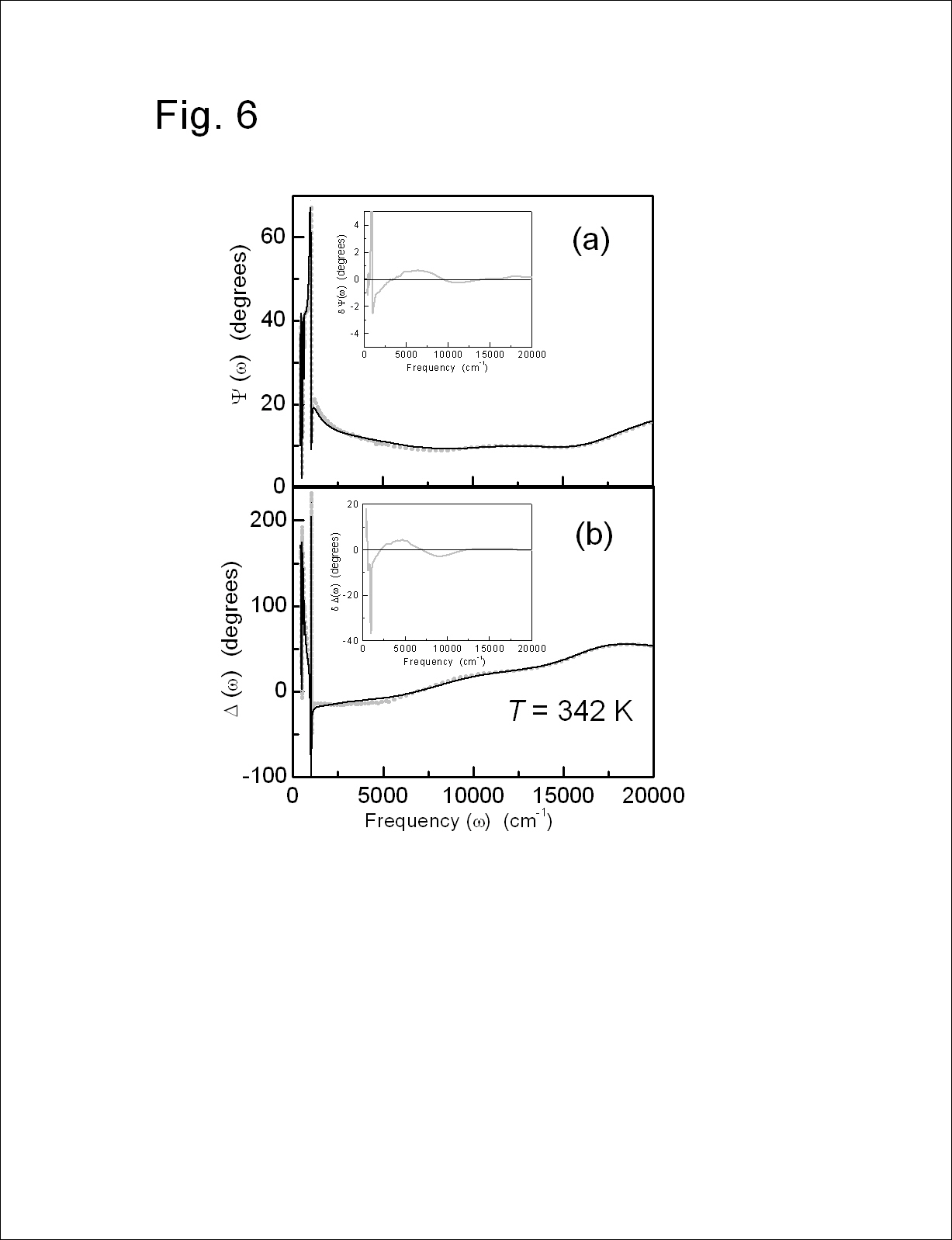}
\end{figure*}

\begin{figure*}[t]
\includegraphics[width=150mm,bb=0 144 620 750,clip]{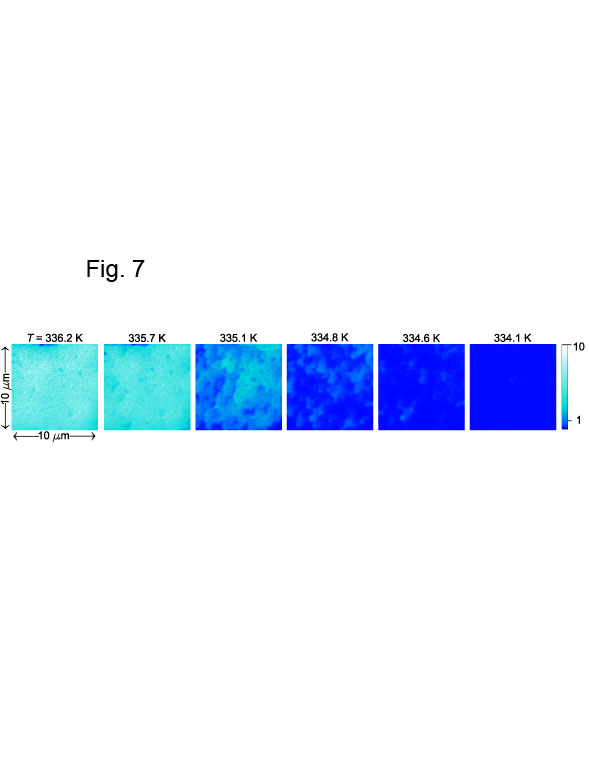}
\end{figure*}

\begin{figure*}[t]
\includegraphics[width=150mm,bb=0 144 620 750,clip]{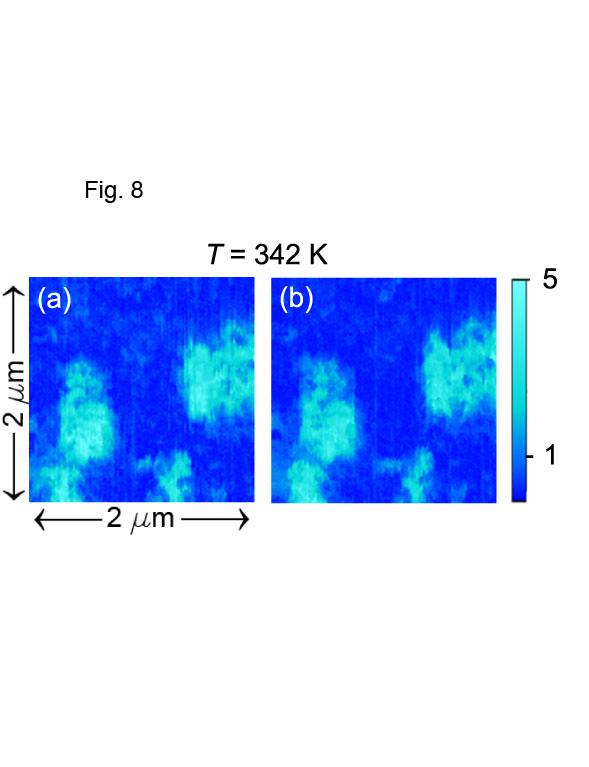}
\end{figure*}

\end{document}